\documentstyle[12pt,aasms]{article}

\begin{document}

\title{A Study of the Wavelength Calibration of {\it NEWSIPS} High-Dispersion
Spectra}

\author{Myron A. Smith$^1$ }
{\footnotesize $^1$STScI/CSC, Space Telescope Science Institute, 3700 
San Martin Dr. Baltimore, MD 21218; ~~Email:~ msmith@stsci.edu 
}

\begin{abstract}

  In this study we cross-correlate many {\it IUE} echellograms of a 
variety of well
observed stars to evaluate systematic error sources in the wavelength 
zero-points (velocities) of all three cameras.
We first evaluate differences between the final archived (``NEWSIPS")
and the originally processed (``IUESIPS") echellograms. These show a
marked time dependence in zero-point for the {\it SWP} camera due to several
revisions of wavelength calibration coefficients used for {\it IUESIPS}. 
Smaller offsets are present for the {\it LWR} camera between the two 
processings. We also evaluated small-amplitude fluctuations in the 
zero-points of the {\it NEWSIPS} wavelength calibration spectra themselves. 
In the case of the {\it SWP} camera, these variations are too complicated
to have been completely removed in the {\it NEWSIPS} wavelength calibration. 
We also examine wavelength zero-point disparities between data obtained both
through the small and large entrance apertures as well as for observations
made by different target acquisitions of faint and bright stars. We also 
find that statistical differences between these alternative observing modes 
are virtually nil. For large-aperture
observations the dominant error source is the placement of the target 
in the aperture. These give rise to non-gaussian, extended ``tails" in 
apparent velocity. We also searched for spurious trends with time. 
Except for a possible trend for faint objects with {\it SWP} camera data, 
we can not detect significant dependences with time. Additionally, we 
discovered a trend with 
telescope focus for datasets derived from intensive monitoring campaigns 
of bright stars. These exhibit a repeatable, one-day ``radial velocity
variation" with a semi-amplitude of nearly 3 km~s$^{-1}$. This pattern appears 
to be a by-product of fluctuations in telescope focus caused by operational
procedures to maintain the ambient instrument temperature.

  In the second part of the paper, we measure the mean zero-point errors of
{\it NEWSIPS} echellogram data with respect to laboratory results by using
the {\it Goddard High Resolution Spectrograph} ({\it GHRS}) spectral atlas
of the O9~V spectral standard 10~Lacertae as an intermediary reference. 
We find that the derived apparent velocity difference for this star is 
essentially zero: -1 ${\pm}$3.5 km~s$^{-1}$. 
Several less precise comparisons lead to similar results, including 
cross-correlations of: (1) spectra of 10 Lac and two stars with similar 
spectra, HD~93521 and HD~60753, (2) lines in common with the {\it SWP} 
camera and {\it GHRS} and {\it STIS} atlases of Arcturus and Procyon, and 
(3) interstellar lines in the {\it GHRS} spectrum of the white dwarf G191-B2B. 
The zero-points of the {\it NEWSIPS}-processed long-wavelength cameras are 
evaluated and are also found to be nearly zero (${\pm}$5 km~s$^{-1}$) 
relative to the Arcturus and Procyon atlas calibrations and relative to one
another. In general, these results do not support the suggestion by 
Gonz\'alez-Riestra that corrections should be introduced to the wavelength 
scales of various {\it NEWSIPS} high-dispersion data products. 
Despite our optimistic results, 
it is obvious that using small {\it IUE} datasets from large-aperture 
observations of arbitrarily chosen stars can contain velocity errors of 
at least a few km~s$^{-1}$.

\end{abstract}

\section{Introduction}
\label{intro}

   During its operational lifetime (26 January 1978 to 30 September 1996) the 
{\it International Ultraviolet Explorer} satellite observed over 3700 objects 
outside the Solar System in its high dispersion mode. Because a high
percentage of these objects have been regularly observed by other satellites
and from the ground, a comparison of radial velocities obtained from this 
instrument is of considerable value to multi-wavelength and time-monitoring 
investigations that utilize {\it IUE} data. However stable a 
spectrograph might be, its calibration is 
susceptible to systematic wavelengths errors, usually arising from such
real-world complications as the differing paths of the stellar and 
emission-line calibration beams through the spectrograph, dispersion 
nonlinearities, and in the case of the {\it IUE}, both image placement in 
the aperture and a wandering of its echelle spectral format on the detector 
surface. In this paper one of our goals will be to examine several 
potential sources of systematic error as a function of various instrumental 
parameters by means of standard cross-correlation tools. Our purpose is to 
give a general evaluation of the robustness of the wavelength calibration of 
the {\it ``NEWSIPS"}\footnote{NEW Spectral Imaging Processing System (see 
Garhart et al. 1997).} which, in contrast to its predecessor ``IUESIPS" 
software, is not known to produce wavelength errors as a function of time 
or other relevant variables.

  Unlike other wavelength comparisons of the {\it IUE} calibration, 
our study relies on cross-correlations of like spectra, that is, of the 
same star. Cross-correlation studies have a high internal accuracy and 
offer safeguards against systematic errors in the measurement process. 
However, they also have the disadvantage of referring the measurements 
to a reference spectrum template for which the zero-point itself must be
determined. Thus, a second goal will be to evaluate the ``absolute" 
wavelength zero-point errors of the {\it IUE} cameras in high-dispersion. 
This assessment has been carried out with respect to instruments on other 
space-borne platforms, specifically the {\it Goddard 
High Resolution Spectrograph; (GHRS}; operational period 1990--1997) and 
the {\it Space Telescope Imaging Spectrograph (STIS}; 1997--present) on 
board the {\it Hubble Space Telescope}. We will also compare zero-points
of the {\it IUE/SWP} wavelength ``system" with 
ground-based measurements of reference stars and absolute (laboratory) 
values compiled in the literature (Brandt et al. 1998).

A list of actual camera sequence numbers for observations used in this
work are available by request from the {\it MAST}.\footnote{Multi-Mission 
Archive at the Space Telescope Science Institute, funded through a NASA 
contract to the STScI.} The spectra utilized were the ``absolute
calibrated" flux extensions of the ``MXHI" FITS files.

\vskip .4in

\section {Procedures and Calibrations} 

\subsection{Cross-Correlation Tools }
\label{xcr}

  A cross-correlation analysis offers both a convenient and efficient 
method of determining wavelength shifts for a series of test spectra with 
respect to a reference spectrum. This is largely because it utilizes the
aggregate signal contained in the spectrum. No pre-conditions are 
required in terms of the shapes or positions of individual spectral features. 
Thus, the presence of practical problems such as line blending or continuum 
placement do not degrade the accuracy of the solution 
as long as these as errors are independent of wavelength.
Even a limited signal-to-noise ratio will decrease the precision of the result, 
but it does not introduce a bias. To implement the cross-correlation analysis 
we used the {\it IUEDAC}
\footnote{IUE Data Analysis Center (IUEDAC) was a group in the {\it IUE}
Project dedicated to writing specialized IDL software for 
the analysis of IUE spectra.} IDL routine {\it CRSCOR}. This program is 
itself a ``wrapper" for a small IDL collection of programs. It computes the 
cross (auto) correlation function according to its definition: that is, by 
shifting iteratively a test spectrum with respect to the reference by a 
series of regularly spaced wavelength increments and computing the sum of 
the product of the flux differences of the shifted and unshifted spectrum 
at each pixel within a specified wavelength or velocity range. This maximum
of this function is located by least-squares polynomial fitting. For
echelle data the natural unit is velocity, and so the pixel
shift produced by the program is converted to km~s$^{-1}$ as an output option.
As implemented, {\it CRSCOR} truncates the start and ending wavelengths of
the spectral array to the same values, thereby avoiding the generation of
any false ``noise" due to wrap-around of shifted arrays. 
The program interpolates the test 
spectrum to the wavelength grid of the reference template to accommodate 
subpixel wavelength shifts. This step also 
facilitates comparisons with spectra from other instruments. To test these 
interpolations we cross-correlated several test spectra against copies of
themselves by using a similar (but not identical) set of wavelengths 
from other observations of the same target.
This comparison tests the accuracy of interpolating spectra
observed at discrete wavelengths. We found that interpolations to one set 
of wavelengths produce false shifts of always less than $\pm{1.0}$ km~s$^{-1}$.

  Our procedure was to run {\it CRSCOR} iteratively for each echelle 
order and to evaluate the net shift of the order as a whole wavelength 
segment. In our procedures we took care to screen the data for
known pathologies of the {\it IUE} cameras. For example, fluxes
near the edges of the order (particularly the long-wavelength end for 
the {\it SWP} camera) were avoided because the noise is high and also because 
systematic errors can be large in the blaze ``ripple correction" 
(Gon\'zalez-Riestra et al. 1998). Such errors often mean sloped continua,
which can lead to a small bias in line centroid positions.
We also excluded small groups of pixels with negative data quality flags 
(usually associated with instrumental ``reseaux") because their fluxes 
are generally meaningless. In these cases, we interpolated the fluxes 
from neighboring pixels. This step has the effect of adding incrementally
to random noise but does not introduce a random shift unless the interpolated
regions in the two spectra being compared are extensive and have different
slopes. Apart from specific tests on zero-point shifts of 
interstellar line systems ($\S$\ref{part2}) discussed below, 
we did not exclude interstellar lines from our cross-correlations. 
Although these lines are formed in a velocity frame shifted from the
photospheric lines, the resulting wavelength shift can be assumed to be 
the same for different instruments because the cross-correlation routine 
does not distinguish between features formed in various velocity frames.
For example, we did not have to exclude ``wind" lines because the shifts 
from orders including these lines did not differ noticeably from shifts of 
adjacent orders. 
  
 In comparing {\it IUE} spectra with spectra from the {\it HST} {\it GHRS} 
and {\it STIS} instruments, we convolved the {\it HST} data by gaussian
broadenings to make their resolutions equivalent to {\it IUE} spectra.
When all these steps were carried out, we experimented with 
{\it CRSCOR} in an interactive mode for the investigations discussed
in $\S$\ref{part2}. After finding that shifts arising from unflagged data
pathologies are rare, we automated the program in order to compare large 
groups of spectra efficiently.

\subsection{The Calibration of Wavelengths in NEWSIPS }
\label{calib}

  Before an analysis of the zero-point errors of wavelengths in the {\it
IUE} cameras can proceed, it is necessary to review the steps taken in
the wavelength calibration of {\it NEWSIPS} data; some of the
following details have not been published in any form.
The calibration of wavelengths in an echelle system requires the prior 
mapping of pixels from a two-dimensional detector surface to a series of 
one-dimensional arrays each of which is represented as a smoothly varying
function of wavelength.
In the raw detector geometry the echelle orders fall at unique angles 
relative to the electronic scan axis, and thus together they run 
nonparallel to one other. In addition, in certain places on the detector 
the spectral format can be displaced by local shears from slight 
nonuniformities within the optical coupling elements.
All of these factors necessitate a detailed geometrical resampling of 
the raw echellogram image to a new surface in which the 
orders run parallel along a common dispersion axis and in which their 
fluxes have been reevaluated at a constant velocity increment.  
{\it NEWSIPS} maps the raw positions to a rectilinear geometry, thereby
simplying the extraction of spectral fluxes 
(for details see Chapters 7--9 of Garhart et al. 1997). Wavelength 
calibration (``WAVECAL") observations of the three cameras were made 
monthly during the mission through the small aperture by means of on-board
platinum-neon (Pt-Ne) lamp source. Early in the mission a tungsten flood 
lamp (``TFLOOD") exposure was added to the emission-line exposure in order 
to enable process software to determine the positions of reseau more
easily. In April, 1990 this practice ceased when studies within the {\it IUE} 
Project demonstrated that these pre-exposures added a noise component 
to the signal which effectively reduced the instrumental dynamic range 
and thus the number of measurable emission lines. 

  With the linearized fluxes extracted, the wavelength system of each 
spectrograph camera could be calibrated by means of a reference calibration 
spectrum by remapping the $(x,y)$ positions of its emission lines from the
original geometry to the rectilinear detector surface. For this step one
may think of the dispersion axis as ``$x$" and of the spatial (or echelle
order $m$) axis as ``$y$." 
Wavelengths of the emission lines were taken from the study of Reader et 
al. (1990), which was commissioned for the purpose of calibrating 
wavelengths of the {\it HST} spectrographs. 
The first of four steps in the {\it NEWSIPS} wavelength calibration 
procedure was interactive and consisted 
of running the {\it IRAF}\footnote{``{\it IRAF}" is distributed 
by the National Optical Astronomy Observatories, which are operated by the 
Association of Universities for Research in Astronomy, Inc., under cooperative 
agreement with the National Science Foundation. } routine {\it ecidentify} to 
map the positions of a set of lines to wavelengths for a reference {\it 
WAVECAL} echellogram.\footnote{The reference echellograms used in this step 
were SWP~38001, LWP~15422, and LWR~18381.}
In this step a quadratic Cheybshev polynomial functon was used to map the 
$(x,y)$ pixel locations in the reference echellogram to a wavelength table.
The second step was to utilize the {\it IRAF} routine
{\it ecreidentify} 
Chebyshev solution determined for the reference echellogram in to
determine predicted pixel positions for a few hundred lines in all
the {\it WAVECAL} observations for a given camera. 
A polynomial fit was performed for each of the echellograms, 
resulting in a set of wavelength zero-points and mean pixel-to-pixel 
wavelength increments for each order. In practice, the wavelength 
zero-points for each ``science echellogram" were fit to a cubic polynomial 
in both time and camera-head temperature (``THDA"). These fits were adopted 
for all science data but of course not for the {\it WAVECAL} calibration 
observations themselves. This last step compensated for image shifts 
statistically by means of a previously derived least squares correlation
of wavelength shift as a function of time and temperature. Note that the
wavelength calibration of the {\it WAVECAL}s technically applies to 
observations made through the small aperture. However, this calibration can
be readily applied to large-aperture observations from prior measurements 
of the offset between the two apertures in the dispersion direction.

  We should pause to point out that whereas the wavelength calibration for
{\it NEWSIPS} was performed with a simultaneous polynomial fit in both the
wavelength and echelle order dimensions, the number of {\it WAVECAL}
emission lines was sufficient to permit a calibration of wavelengths for
most echelle orders separately (Smith 1991). The first (``global")
approach is clearly the preferred one if pixel-to-pixel spacings
on the detector are regular because a large number of lines
can contribute to the solution for any individual order. However, during 
the {\it IUE} lifetime the cameras suffered both electro-optical 
distortions and differential shifts arising from shears between 
fiber-optical bundles within the coupling-plate of the camera. A global 
wavelength solution tends to smooth over these small-scale distortions,
causing correlated order-to-order errors in derived wavelengths. 
Consequently, with exceptions noted by Garhart et al. (1997), the {\it 
ecidentify} steps described above were implemented for individual spectral 
orders in the {\it NEWSIPS} calibration.

\subsection{ Cross-Correlation of Wavelength Calibration Spectra }
\label{xccalib}

  In this section we discuss various time-dependent characteristics of the 
{\it WAVECAL} echellograms used to calibrate the science data. It should
be emphasized that most of the following results were initially derived for 
use in removing low-order time-trends in the science data within the
pipeline {\it NEWSIPS} processing of science spectra.
The characteristics for {\it WAVECAL} spectra discussed below 
have only an indirect effect on the systematics and accuracy of the 
zero-points of science observations.

  Starting with the cross-correlation software described in $\S$\ref{xcr}, 
we compared wavelengths for some 50 echelle orders in all small-aperture, 
optimally exposed {\it WAVECAL} spectra (278 for LWP, 291 for LWR, and 
468 for SWP) against the reference spectra used by Garhart et al. (1997) 
in the {\it ecreidentify} step of their calibration. The cross-correlations
showed that the zero-point shifts found for the {\it SWP} data have the 
most complicated dependences of all three cameras and, in particular, that 
they exhibit fluctuations in wavelength as well as time. 
An example of the dissimilar
dependences with time at different wavelengths for this camera is 
depicted in Figure~\ref{ms1fig1} and requires some discussion. 
In analyzing the {\it WAVECAL} data, we
found that it was necessary to discard the pre-1979.0 (satellite
commissioning period) {\it SWP} camera {\it WAVECAL}s
from further analysis because the zero-points otherwise showed
a very steep increase during 1978. We then fit the shifts for the 
remaining sample of 415 spectra with 7th-degree polynomial both 
in time and echelle order. No such complications were found for the
long-wavelength cameras. The cross-correlation results for all three
cameras are plotted with time 
in Figures~\ref{ms1fig2} and \ref{ms1fig3}. 
Figure~\ref{ms1fig2} was constructed by averaging the wavelength shifts
such as those shown in Fig.~\ref{ms1fig1} over echelle orders. 

 The complicated dependence of the {\it SWP} camera wavelength zero-points 
as a function of time and spatial direction (echelle order) is reminiscent 
of the complicated decay of this camera's ``null flux" surface with time, as 
recently documented by Gonz\'alez-Riestra (1998) and Smith (1999). 
During 1979--80 the zero-point of the {\it WAVECAL} spectra increased 
smoothly by $\approx$+15 km~s$^{-1}$ (about 2 pixels) for order $m$ = 70 
($\lambda$1170), decreased by about -4 km~s$^{-1}$ during the mid-1980's, 
and decreased rapidly during the last 2--3 years of the mission.
For $m$ = 118 ($\lambda$1970), at the other end of the camera, the
corresponding changes in zero-point are +4.5 km~s$^{-1}$, 0 km~s$^{-1}$, 
and -2 km~s$^{-1}$. One sees that the greatest changes occur in the 
``short-wavelength" corner of the echelle surface, primarily during both
the early and late stages of the {\it IUE} mission. Since changes occurred
in the raw background flux at about the same epochs and in the same
region of the camera as the changes we report here,
it is possible that changes in the camera characteristics influenced the 
positions of the echelle format as well as the net fluxes of these orders
as a function of time. 

 In contrast, the {\it WAVECAL} zero-points for the long-wavelength cameras 
can be fit with low order polynomials, and they do not have a marked
dependence on position on the detector. The {\it LWP} camera
is particularly well behaved in this respect since its wavelengths
(Fig.~\ref{ms1fig2}a) change linearly by 
+3 km~s$^{-1}$ over the interval 1980--1996.8. For the {\it LWR} camera 
(Fig.~\ref{ms1fig2}b) the wavelengths change by about +4 km~s$^{-1}$ 
during 1978--1980 and remain almost constant thereafter. 
  
  Low-order (cubic) fits for the zero-point dependences of the {\it
WAVECAL}s for the three cameras were hard-coded 
into {\it NEWSIPS} in order to remove such trends from the science data. 
For the {\it LWP} camera data the dependence on 
time is linear, so the trends could be removed completely. For the {\it 
LWR} camera the removal of trends is probably very good except for fitting
the rapid increase during the first 2--3 years of the mission. 
A cubic polynomial cannot fully remove this early-epoch trend. 
As implied above, the complex dependence for the {\it SWP} camera
cannot be accurately fit with a cubic polynomial function, particularly 
either early or late in the mission. This can be seen by comparing
the cubic fit to the shifts in Figure~\ref{ms1fig2}. The differences between
this line and the zero-points of the individual {\it WAVECAL} spectra imply 
that, independent of other error sources, {\it SWP} data can be expected 
to have small ($\le$ $\pm{2}$ km~s$^{-1}$) epoch-dependent errors. 

  Aside from temporal trends in the mean shifts, one can see from
Figures~\ref{ms1fig2} and \ref{ms1fig3} that the r.m.s. scatter 
characteristics of zero-points among {\it WAVECAL} spectra also 
change with time. Considering the {\it SWP} camera results first,
Fig.~\ref{ms1fig2} shows that starting sometime in 1988--90 the scatter 
decreased to only about 60\% of its initial value of $\pm{2.5}$ km~s$^{-1}$. 
Also, the statistical outlying points (defined as those differing from
the epochal mean by at least 1 pixel or 8~km$^{-1}$) decreased in 
occurrence from over 10\% of all obervations to only about 4\%. 
It is possible that these changes are caused by the termination of
adding {\it TFLOOD} lamp exposures onto the {\it WAVECAL} 
spectra. Other changes were made during this period, such as taking 
these observations under more tightly defined constraints in THDA and focus,
might have contributed to the improved scatter characteristics as well.
The scatter for {\it LWR WAVECAL} data seems to decrease well before 1990, 
so the practice of adding {\it TFLOOD} does not seem to be 
important. Also, notice that the scatter for the 
{\it LWP} camera actually seems to have {\em increased} markedly in 1990. 
The suddenness of this change, particularly in the {\it LWP} camera, 
suggests that the change in adding {\it TFLOOD} flux {\em might} have 
contributed to the scatter of the {\it WAVECAL} zero-points for the
calibration of this camera.

\section{Systematics in IUE Parameter Space }
\label{part1}

\subsection{Comparison of IUESIPS and NEWSIPS}  

\subsubsection{Short-Wavelength Prime Camera:}
\label{iuenew}

  There is no doubt that for radial velocity or wavelength studies involving 
{\it IUE} data, the {\it NEWSIPS} processing is much superior to that of 
{\it IUESIPS}. The wavelength calibration coefficients were periodically
updated during the mission in {\it IUESIPS}, so datasets spanning several 
years cannot have consistent wavelengths. 
In this section we quantify the zero-point differences between 
{\it IUESIPS} and {\it NEWSIPS} calibrations during the mission.

  For this purpose we chose hot stars that the {\it IUE} observed over nearly
the entire satellite lifetime. 
The first two stars, 10~Lac (09~V) and $\tau$~Sco 
(B0.2 V), are sharp-lined spectral standards with no known radial velocity 
variations. We obtained all {\it SWP} high-dispersion observations from the 
{\it MAST} database after 1979.0 (except for SWP~01205, 01765, \& 02051 for 
which {\it IUESIPS} data were not available) and cross-correlated pairs of 
{\it IUESIPS} and {\it NEWSIPS} fluxes in spectrum segments in orders $m$ = 
50--119. These results are shown in Figure~\ref{ms1fig4}. 
The series of dotted vertical lines 
mark the dates in which then newly-derived wavelength calibrations 
were implemented into {\it IUESIPS} processing.
We also cross-correlated 33 echellograms from both processing systems 
for six white dwarfs which were analyzed by 
Holberg, Barstow, \& Sion (1998; hereafter ``HBS").
These stars are WD 0005+511, WD0044-121, WD0621-376, EG-102, GD-394, 
and WD2211-495, and in all cases their spectral lines are sparse and broad. 
Thus, we found it advantageous to cross-correlate the same 50 orders 
of these spectra as those utilized for our study of
10~Lac and $\tau$~Sco. We also cross-correlated a few orders 
dominated by interstellar lines. 
We found a mean zero-point difference, {\it IUESIPS} - {\it NEWSIPS}, 
among the sample of HSB white dwarfs of +9.5 $\pm{1}$ km~s$^{-1}$.
This compares rather well with the difference of
+9.1 km~s$^{-1}$ given by HSB (referred to $\lambda$1400). 

  The dependences of zero-point with time for both white dwarfs and hot
main sequence stars in Fig.~\ref{ms1fig4} are likewise similar to the 
HBS results. Thus, the form of this zeropoint-difference curve is 
insensitive to the sample of stars chosen. Rather, it is largely determined 
by the dates of the revisions of the {\it IUESIPS} calibration constants 
were implemented (see vertical dotted lines in the figure). 
Notice that the net difference, {\it IUESIPS - NEWSIPS}, 
is positive for all epochs and increases 
during the late-1980's to a rather well determined maximum of +18$\pm{1}$ 
km~s$^{-1}$ at 1992.5. It then decreases quickly during 1993 and is 
perhaps constant thereafter. The {\it IUESIPS} wavelength calibration was 
unsettled prior to 1983, producing a large scatter compared to 
the homogeneously-calibrated {\it NEWSIPS} system.

\subsubsection{ Long-Wavelength Cameras:}

  We also compared {\it IUESIPS}--{\it NEWSIPS} wavelength zero-point 
differences for the two long-wavelength cameras; we believe this is the 
first such study of its kind. This endeavor turned out to be far more 
difficult because of the paucity of data over an extended time for any given 
star. As a result, we were unable to extract information on the time evolution 
of the wavelength zero-point differences for these two cameras. 

  To determine a {\it IUESIPS} - {\it NEWSIPS} mean difference for the
two long-wavelength cameras, 
we considered the K2~III star $\alpha$~Boo as well as
three of the four photometric calibration (``{\it PHCAL}") standards 
for the {\it IUE} mission, $\zeta$~Cas, $\tau$~Sco, and $\eta$~UMa. 
A number of echelle orders of spectra of the hot stars do not 
contain many lines in this spectral region, 
so we selected several orders interactively that gave a reasonable signal
in the cross-correlation function and used them alone.
The averages of the {\it IUESIPS} - {\it NEWSIPS} differences can be
summarized as follows. For the {\it LWP} camera invesigation
we obtained +0.5 km~s$^{-1}$ for $\zeta$~Cas (126 spectra), +3.2 
km~s$^{-1}$ for $\tau$~Sco (91 spectra), +3.6 km~s$^{-1}$ for $\eta$~UMa 
(165 spectra), and 4.6 km~s$^{-1}$ for $\alpha$~Boo (19 spectra). 
For the {\it LWR} camera the corresponding differences are: 
+5.9 km~s$^{-1}$ for $\zeta$~Cas (65 spectra), +6.9 km~s$^{-1}$ for 
$\tau$~Sco (38 spectra), +6.5 km~s$^{-1}$ for $\eta$~UMa (107 spectra), 
and 6.3 km~s$^{-1}$ for $\alpha$~Boo (11 spectra). Taking the 
straight mean of these results, we have the following differences:

\centerline{$\Delta$RV$_{LWP}$ ~=~ RV$_{IUESIPS}$ ~--~ RV$_{NEWSIPS}$ ~=~ +3.0 
$\pm{2}$ km~s$^{-1}$ }
\centerline{$\Delta$RV$_{LWR}$ ~=~ RV$_{IUESIPS}$ ~--~ RV$_{NEWSIPS}$ ~=~ +6.4 
$\pm{0.5}$ km~s$^{-1}$ }

\noindent In these relations we have applied the (larger) r.m.s. values from
the {\it LWP} camera to the {\it LWR} camera. 
The shift for the {\it LWR} camera is almost certainly significantly 
different from zero. As with the {\it SWP} camera, the difference(s)
are likely to arise from periodic changes in the 
{\it IUESIPS} wavelength calibration.

\subsection{Zero-Points Errors between Large and Small Apertures}
\label{aperr}

  Each of the three {\it IUE} cameras included a pair of science apertures 
in the telescope plane, a small (``{\it SMAP}") and large (``{\it LGAP}") 
aperture.
The SMAP was used primarily during the first few years of the mission because 
of concerns about the wavelength stability for large-aperture observations.
Thus, acquisition/guiding errors might result in light losses, but they would 
not so seriously affect wavelength accuracy as they would for the 
{\it LGAP}. With experience in acquiring objects in the {\it Fine Error 
Sensor} ({\it FES}), the {\it IUE} Project realized that
better wavelength precisions could be obtained than initially anticipated,
and observations through the large aperture eventually became routine. 
A related operational issue was the different centroiding modes in which the 
{\it FES} was used to acquire targets prior to guiding. In particular, the
so-called ``underlap" and ``overlap" modes were to acquire bright and faint 
stars, respectively. Although the underlap mode in principle causes an offset
in the target position, pains were taken to correct for this offset.
To optimize these complimentary acquisition strategies, 
the Project settled on a suitable magnitude demarcation of m$_{v}$ = 4.5. 
Thus, the different acquisition strategies can lead potentially to differences
in the wavelength scale between bright and faint stars. 

   To evaluate aperture-to-aperture differences, we proceeded 
to evaluate shifts from cross-correlations of similar spectra. 
We chose {\it NEWSIPS}-processed spectra from several well 
observed OB and white dwarf stars. Table~\ref{tab1} 
gives the numbers of {\it IUE} high-dispersion echellograms used 
for the analysis of the {\it SWP} camera data. As before, these data were 
cross-correlated for echelle orders $m$ = 70--119. The zero-point errors
are expressed in velocity units, a represention that we justify in 
$\S$\ref{wavsys} Table~\ref{tab1} shows that both 
the mean differences, expressed as RV$_{LGAP}$ - RV$_{SMAP}$ velocities, are 
-0.8 km~s$^{-1}$ for six bright stars and -0.7 km~s$^{-1}$ for six white 
dwarfs. The mean shift for any one of these stars is accurate to about $\pm{2}$ 
km~s$^{-1}$ or better. Thus, 
there are no noticeable systematic differences for shifts between the two 
apertures and between the over-/underlap modes guiding modes. 

  Table~\ref{tab1} 
lists similar results for three stars for the {\it LWP} camera and four
stars for the {\it LWR}. The mean {\it LGAP} - {\it SMAP} difference
for the two cameras are
-1.2$\pm{3}$ and +0.2$\pm{3}$ km~s$^{-1}$, respectively. The quoted errors
are sums (added in quadrature) of both {\it LGAP} - {\it SMAP} differences 
and zero-points of individual {\it SMAP} observations. From this work 
it is clear that no major systematic aperture-to-aperture differences are 
detectable for any of the three cameras.

\subsection{Dependence of Zero-Point with Time }
\label{timetr}

  One may use the large {\it SWP} camera datasets just discussed to
determine if trends are present in science observations with respect to time. 
Table~\ref{tab2} exhibits the linear regression slopes with time for the 
large-aperture data for the three cameras. The data represented are in most
cases the same as in Table~\ref{tab1}. 
The second row of each data group lists the ratio of the regression slope
to the r.m.s. error $\sigma$ in the slope. For the {\it SWP} data the
regressions for three stars are significant to at least three sigma. 
For one of these three stars, $\zeta$~Oph, the non-zero slope was caused 
mainly from data by an intensive campaign conducted by one observer. 
Investigation showed that these differences could be traced to use of
a different reference position in the {\it FES}. 
Thus, the derived slope for this star is questionable. 

 Unlike the case of $\zeta$~Oph, the {\it SWP} data for two well-observed 
stars, HD~60753 and BD+75$^{o}$325 in Table~\ref{tab2}, show clear negative 
trends with time (see Figure~\ref{ms1fig5}) that cannot be readily attributed 
to observational practices or instrumental conditions. 
A similar trend may also be present in the data of the A5~V star HD~11636 
(not shown), for which the far-UV fluxes are faint. 
We might speculate that this apparent trend is actually the norm for UV-faint 
sources. For example, it is possible that these trends, if real, are 
by-products of an unexplained source of scattered light, i.e., ``the streak,"
which appeared in 1991.
In fact, the scattered light from this artifact introduced a gradient  
in the light admitted across both the small and large apertures required 
that {\it IUE} telescope operators to adopt a new acquisition reference point 
in the {\it FES} to avoid shifts in the spectra of several km~s$^{-1}$ 
(Pitts 2000). This displacement is similar to the change in
zero-points documented for HD~60753 and BD+75$^{o}$325. 
Overall, we find that all but one of the 11 stars in the sample represented 
in Table~\ref{tab2} shows a negative slope. 
When subjected to a statistical ``Sign Test" (Dixon and Massey 1951)," 
the signs of these slopes are weakly significant at the 2$\sigma$ level. 
We conclude that over a 15 year interval a trend of
several km~s$^{-1}$ {\em might} exist for faint objects. We stress that this
trend is not apparent in our bright objects results.

   In contrast to the {\it SWP} camera results, we have found no clear 
velocity trends for spectra observed with the {\it LWP} and {\it LWR} camera. 
The {\it LWP} sample consisted of 
observations of eight stars: $\tau$~Sco, 10~Lac, $\eta$ UMa, $\zeta$~Cas, 
HD~93521, RR~Tel,\footnote{The UV spectrum of RR~Tel, a symbiotic binary,
is dominated by several strong emission lines. The cross-correlation results 
we report are for those orders containing these emission lines. Since these
orders contain only 1-2 strong lines apiece the r.m.s. errors are large,
$\pm{6.5}$ km~s$^{-1}$. Other orders in these spectra contain faint absorption
lines, which produce an r.m.s. in their shifts of nearly the same value, 
$\pm{7.1}$ km~s$^{-1}$. The orders containing emission lines give
comparable quality cross-correlation results as those without them. RR~Tel 
shows by far the poorest cross-correlations of any spectra in our sample. 
}
BD+28$^{o}$4211, and BD+75$^{o}$325. 
The distributions of shifts of all these datasets showed equal numbers of 
positive and negative slopes with time, none of which
were significant above the level of 1.5$\sigma$. 
The occurrence in October 1983 of a large ``flare," 
a developing hole in the high voltage assembly, thereby producing a high, 
localized background, complicated the analysis of late-epoch data
for this camera. The flare caused the {\it IUE} Project to reduce the camera
voltage, resulting in a loss of sensitivity. This sequence of events led
to a curtailment in the camera's use and an abbreviated time interval
used to monitor instrumental trends of well observed objects.
Among our sample of program bright stars and white dwarfs, we were able to 
find only two stars, $\zeta$~Cas and $\tau$~Sco, with an appreciable number 
of same-aperture observations before and after the flare. Of these two 
targets, $\zeta$~Cas is the better sampled in time. Its cross-correlation 
shifts exhibited no trend with time, but they did exhibit a pronounced 
increase in scatter (about a factor of two) after 1984.0. The data for
$\tau$~Sco showed an apparent slope of -1 km~s$^{-1}$~yr$^{-1}$, but this 
could be an artifact of a greatly increased scatter after 1984.0.
An examination of the shifts as a function of wavelength for this star 
revealed no wavelength-dependence. This would be surprising if the flare 
were the cause of wavelength shifts. Our conclusion from this investigation 
is that the wavelength zero-points of {\it LWR} observations taken after 
the 1983 flare event are less reliable than before. However, this may not 
be a direct consequence of the flare itself. Rather, we suspect that the 
increased scatter arises from the decreased usage of the camera and the 
resulting lower accuracy of the necessary redetermination of
positions in the {\it FES} used to acquire an object.

\subsection{Dependence with Focus and Temperature }

  Early in the {\it IUE} mission it was discovered that the camera image 
format shifts frequently in both the spectral and spatial directions with time 
and camera temperature. Camera temperature was measured roughly by thermistors
located in any of several areas of the camera head amplifier (not an
ideal location to measure the temperature of the spectrograph bench itself),
and one of these was selected to provide a consistent refererence. This 
provided a ``THDA" index which could be used to calibrate wavelength 
zero-point shifts and apply them to the processing of a particular 
observation in {\it IUESIPS} and {\it NEWSIPS}. 
Motion of the spectral format on the detector was generally caused by 
telescope flexures and, under routine operating conditions, secondarily 
by changes in electro-optical properties of the camera. Thus, fluctuations  
in satellite temperature (which was primarily a function of 
the telescope's orientation angle with respect to the Sun) caused the 
telescope to expand or contract and the focus to change. 
The focus was usually controlled indirectly by turning on or off 
a heater located behind the primary mirror. An alternate control 
procedure consisted of toggling a heater mounted on the spectrograph deck. 
Although the camera temperature and detector image shifts were correlated, 
a hysteresis arising from the instrument's thermal history tended to
weaken the efficacy of controlling the focus by commands to the spacecraft.
This control was weakened further by other environmental factors, such as 
the Earth's eccentric orbit around the Sun and earthglow. Consequently, the 
{\it IUE} Operations team developed a conservative strategy of maintaining 
the THDA temperature within certain limits. During the course of the {\it IUE} 
mission statistical correlations were derived from shifts of {\it WAVECAL} 
spectra as a function of the {\it THDA} and focus parameters. 
These shifts were parameterized
by means of a polynomial fit, so {\it NEWSIPS} could remove them to first 
order in its assignment of wavelengths. 

  As part of our cross-correlation analysis, we searched for 
a correlation between zero-point shifts and {\it THDA} and focus
values at the time of an observation. Generally, 
we were unable to find a convincing correlation. 
Except for an accident of an unrelated contemporaneous research program, 
this search might have stopped with this null result. However, while 
investigating spectral variations in B stars for another research program, 
we used the cross-correlation tools from the present study to shift spectra 
to the same wavelengths before manipulating the data further. In doing so,
we noticed a time-dependent cycle of about a day (see Figure~\ref{ms1fig6}) 
in the results for a time series of 22 observations conducted in 1995 on the
B3Ve star $\alpha$~Eri. In searching through the {\it IUE} archival database 
further, we found an extraordinary set of 181 continuous observations on the 
B0.5V star $\epsilon$~Per in 1996 obtained by Dr. D. Gies. 
Figure~\ref{ms1fig7} shows that the same one-day cycle is present in the data 
for this star.
A string of observations of HD~93521 at 1994.2 (not shown) exhibits a 
similar 1-day period and few km~s$^{-1}$ semiamplitude over three days.

  Because the cycle lengths in these shifts are all close to one day, 
we investigated first whether the satellite's orbital motion might have
somehow been neglected in determining wavelengths for an individual 
observation. However, we were able to discount that possibility.

  A more successful attempt to explain the apparent 1-day velocity cycles of 
$\alpha$~Eri, $\epsilon$~Per, and HD~93521 was to investigate the effects
of varying the instrument temperature by adjusting the telescope focus. 
Cycling the camera-deck heater produced correlated responses of the
THDA and telescope-focus values, especially when the 
ambient spacecraft temperature was lower than nominal, as for the 1996 
observations of $\epsilon$~Per. During this particular monitoring series, 
the good correlation between temperature and focus indicates that the focus 
was controlled by the deck heater. In the time series on $\alpha$~Eri the
temperature was not quite so low. Then an adequate means of controlling 
the focus was to cycle the more distant telescope mirror heater. 
Since the telescope heater was isolated from the spectrograph, the locally
measured {\it THDA} value did not correlate well with the focus changes
of the telescope (see lower dashed line in figure). However, in either
case the principle was the same, that heating could be applied to prevent  
the focus from drifting to large negative values. The important point is 
that this application caused an overcorrection of the focus. 
The focus values would then swing (relatively) positive until a low ambient 
temperature again reversed the change of the focus, causing a new 
thermal-focus cycle to ensue. All told, the changes in focus from either 
thermal-control technique caused the velocity to become first too negative 
and then too positive by a few km~s$^{-1}$. Although we have found the 
velocity-focus correlation only in {\it SWP} camera datasets so far, it is
probably present in long-wavelength data as well.

  Figure~\ref{ms1fig8} exhibits the correlation explicitly between 
telescope focus value and zero-point for the 
observing campaign on $\epsilon$~Per. A similar plot can be constructed 
for THDA instead of focus, but the $\alpha$~Eri data imply that the 
correlation of velocity with THDA for the $\epsilon$~Per data is a result 
of temperature excursions changing the focus, and {\em not} a shift caused 
by the temperature variation within the camera. 
Note especially that Fig.~\ref{ms1fig8} 
shows that the correlation arises only in the limited focus range of -2.0 to 
-3.7 (instrumental units). As noted above, we also searched our results for 
a dependence on focus and THDA values in our study of time-dependent 
correlations, but we found none. We suspect the reason is that {\it IUE} 
observations of most objects were obtained at different epochs 
and with different target-centering practices. 
This would tend to conceal any correlation over a small critical subrange 
of focus values in our searches for trends in the stars of Table~\ref{tab1}.
However, we expect the pattern shown in Fig.~\ref{ms1fig8} is present in 
all high-dispersion data to some extent. If so, it is a a secondary source
of radial velocity error.

\section{Radial Velocity Zero-Point Errors}
\label{part2}

\subsection{Wavelength-Systematics? } 
\label{wavsys}

 In our earlier discussion we focused only the temporal results because we 
did not find any systematic dependences on wavelength. In Figure~\ref{ms1fig9} 
we demonstrate this further by exhibiting the averages and confidence belts
($\pm{1\sigma}$) of shifts for 16 stars at each wavelength interval in
our {\it SWP} camera study. The mean r.m.s. over wavelength for all 
stars is $\pm{2.4}$ km~s$^{-1}$ while the adjacent order-to-order shift 
is only $\pm{1.0}$ km~s$^{-1}$. These errors are generally 
wavelength-independent except below $\approx\lambda$1230. 
Below this wavelength the errors increase because the detector loses its 
sensitivity to flux. Because there is no clear trend in wavelength, we will 
refer to the zero-point error in terms of apparent {\em velocity} rather 
than wavelength.

  A comparison of the velocity fluctuations resulting from the 
cross-correlations among different stars suggests that the
precision of a zero-point determination varies as a weak 
power of the reciprocal of the {\em net} flux (detector counts). In addition, 
we find that order-to-order shifts are measurably larger for spectra 
of white dwarfs and rapid rotators (e.g., $\eta$~UMa, $\zeta$~Cas). 
This fact suggests that the random 
cross-correlation errors increase slightly with line broadening, namely 
from about $\pm{3}$ km~s$^{-1}$ (e.g., $\tau$~Sco) to  $\pm{4}$ km~s$^{-1}$ 
(e.g., $\eta$~UMa).

\subsection{SWP Analysis} 
\label{swpan}

\subsubsection{IUE versus GHRS }

\noindent a) Available Short-Wavelength Atlases  
\label{atlas}   

  Since the completion of the processing of {\it IUE} data with NEWSIPS
in 1997, new spectral atlases have been constructed using 
the {\it GHRS} wavelength calibration. Both the {\it IUE/NEWSIPS} and
{\it GHRS} wavelength calibrations are derived from 
a recent homogeneous study of ultraviolet platinum lines
(Reader et al. 1990). The most important atlas for purposes of comparison 
to {\it IUE/SWP} data is the Brandt et al. (1998) atlas of 10~Lacertae. 
This work covers the region $\lambda\lambda$1181--1777, which is most of 
the wavelength range covered by the SWP camera ($\lambda\lambda$1160--1980). 
In order to insure that image wandering could not contribute significantly 
to errors, all observations for this atlas were made through the {\it GHRS} 
Small Science Aperture. 
Aside from being a spectral standard (Walborn 1972), 10~Lac is an ideal
object for calibrating the wavelength scale. 10~Lac is an apparently single 
star, and its spectrum has numerous sharp photospheric lines. Fullerton (1990) 
found no measurable variations over time to a precision of perhaps $\pm$1 
km~s$^{-1}$. In addition to plotted specta, the Brandt et al. atlas gives 
measured wavelengths of over 700 lines, which can be compared 
to published laboratory values.\footnote{The full table of measured and 
laboratory wavelengths used by Brandt et al. is available at the URL 
http://archive.stsci.edu/hst/atlas10lac/datalist.html~.}
This information is adequate to permit the determination of an effective
mean velocity of the {\it IUE} SWP camera relative to  ground-based velocity 
measurements of 10 Lac (Hobbs 1969, Grayzeck \& Kerr 1974, Stokes 1978).

\noindent b) The Procedure

  In this subsection we will link the apparent velocities of {\it SWP} {\it 
LGAP} spectra of 10~Lac to the laboratory system. Because this connection 
takes several steps, we will first outline them and then discuss the actual
cross-correlation results:

\noindent {\em Step 1,~ $\Delta$RV$_{SWP}$ =
RV$_{SWP-LGAP}$ - RV$_{SWP-SMAP}$}: 

\noindent This is a velocity correction for {\it unpredicted} difference
in wavelengths of lines measured in {\it IUE} spectra taken through the 
large aperture relative to the small aperture. Such errors might arise from
potential measurement error between the apertures or (more likely)
from changes in the {\it FES} reference position used for
target acquisition.

\noindent {\em Step 2,~ $\Delta$RV$_{IUE}$ = $\Delta$RV$_{SWP-LGAP}$ - 
RV$_{GHRS}$}: 

This term is the instrument-to-instrument correction for the SWP/SMAP 
wavelength zero-point referred to the zero-point of the {\it GHRS} 
(small-aperture) atlas of 10 Lac.

\noindent {\em Step 3,~ $\Delta$RV$_{UV}$ = $\Delta$RV$_{GHRS}$ -
RV$_{lab}$}:

This correction accounts for any difference between the  
radial velocity derived from differences in the measured ({\it GHRS}) and 
laboratory wavelengths. These could result from an error in the calibration
of the {\it GHRS}. We note in particular that the Brandt et al. wavelengths
of photospheric lines in 10~Lac were measured from mainly iron-group element 
lines whereas the {\em calibration}
of wavelengths for the {\it GHRS} systems was derived from Pt line 
wavelengths. \footnote{The line list for SWP camera data was culled 
to 288 lines from the Reader et al. (1990) list. 
This is about twice the number of lines used for the IUESIPS calibration. 
The line list for high-dispersion LWP/LWR data contains 479 lines.}

\noindent {\em Step 4,~ $\Delta$RV$_{star}$ = RV$_{lab}$} - 
$\Delta$RV$_{optical}$: 

The final step is the correction for the star's adopted heliocentric radial 
velocity from ground-based measurements. The ground-based value can be
compared finally to the velocity obtained from appying Steps~1--3. 
In our chain of comparisons, the defining reference is the radial 
velocity of 10~Lac ( -9.7 km~s$^{-1}$; 
Wilson 1953). The addition of the results from 
Steps \#1--4, 1 + 2 + 3 + 4, gives the net discrepancy between the 
{\it SWP LGAP} and the laboratory wavelengths.


\noindent c) Comparison with 10 Lacertae Atlas
\label{10laccmp}

\noindent {\em Analysis for Step 1,  $\Delta$RV$_{SWP}$ (correction from
   {\it IUE} large to small aperture):}

In $\S$\ref{part1} we showed that {\it SWP} camera data for several
bright hot stars do not have discernible velocity trends with time. 
As Figure~\ref{ms1fig10} shows, the test star, 10~Lac, is no exception. 
The mean aperture-to-aperture difference from all available 10~Lac 
spectra is $\Delta$RV$_{SWP}$ = +0.4 km~s$^{-1}$. 
At +1.2 km~s$^{-1}$ $\pm${0.8} km~s$^{-1}$, this value differs slightly 
from the average velocity discrepancy for 12 stars represented in 
Table~\ref{tab1}. 

  As an important digression, we should add that we have searched 
for nonrandom positionings of the image of 10~Lac 
within the LGAP. Figure~\ref{ms1fig11} exhibits the histogram of the 
SWP/LGAP sample of spectra. The centroid of this distribution is 
arbitrarily set to zero. If one first fits the distribution to a single 
gaussian, the r.m.s. dispersion is 4 km~s$^{-1}$. A more convincing 
representation is a bimodal gaussian (see figure) with $\sigma_{1}$,
$\sigma_{2}$ = 2.8 km~s$^{-1}$ per observation. 
The corresponding errors in the
means of the two peaks are 0.5 km~s$^{-1}$ and $\pm${1.2} km~s$^{-1}$, 
respectively. Investigation of the times of the 
observations shows that most of the secondary peak
arises from spectra taken in 1993 and 1995, when different aperture 
reference points were used to acquire the star. It is possible that
target acquisitions at other reference points may be reason for the 
skewed high-velocity tail of the distribution. 
Departures from gaussian distributions are common among apparent velocities 
found from our large data samples, and the extended negative and positive
tails that they represent in a histogram seem to be present equally.
It may be significant that the distribution of cross-correlation shifts
taken from the 1996 data series for $\epsilon$~Per, with the sinusoidal
component removed, generally follows a gaussian distribution and in
particular has no extended tails. This fact suggests that the effective
position of the star image in the aperture is the primary reason for 
outlying velocities in our cross-correlation results. In this picture 
the initial exposures of a target in a time sequence are likely to exhibit 
a large scatter because the telescope temperature and focus had not yet 
equilibriated. After an hour or two the error distribution of the remaining 
observations became tighter and also more closely resembled a gaussian.

\noindent {\em Analysis for Step 2,  $\Delta$RV$_{IUE}$~
(correction from {\it IUE} to {\it GHRS}):} 

   Cross-correlation shifts were computed as before among 41 spectral 
orders in six {\it SWP SMAP} echellograms against the Brandt et al. atlas of 
10~Lac. These shifts are indicated by the solid line in Figure~\ref{ms1fig12}.
The average, $\Delta$RV$_{IUE}$, is +0.6 km~s$^{-1}$, while the r.m.s. 
error per observation is $\pm{1.5}$ km~s$^{-1}$. This gives an 
error in the mean velocity determination of $\pm{0.2}$ km~s$^{-1}$. 
This value is negligible compared to errors from the other steps.

\noindent {\em Analysis for Step 3,  $\Delta$RV$_{UV}$~ (correction from
{\it GHRS} to laboratory system):}

   Brandt et al. (1998) selected 24 photospheric lines from the total sample 
of 705 lines they identified in their atlas and measured a mean velocity 
difference of -7 km~s$^{-1}$ with respect to laboratory wavelengths in the 
literature. However, these authors did not specify which lines they used.  
We have compared the wavelengths of the entire dataset and find instead a 
mean velocity difference of -11.6 km~s$^{-1}$. We repeated this comparison by 
taking a number of subsamples of lines. These differences were found to lie 
always within $\pm{2}$ km~s$^{-1}$ of the latter value. We adopt this value
for $\Delta$RV$_{UV}$.

\indent {\em Analysis for Step 4,  $\Delta$RV$_{star}$~ (correction for
10 Lac's heliocentric radial velocity): }

 The optical catalog radial velocity value for 10~Lac is -9.7 km~s$^{-1}$ and
requires no discussion. The sign of this radial velocity will be reversed
below to refer the velocity system to the {\it GHRS} scale. From the 
agreement of values for this star from different instruments in the 
literature, a reasonable error for the optical determination is 
$\pm{1}$ km~s$^{-1}$.

   In summary, the combination of the four velocities gives a net error
of = +0.4 + 0.6 -11.6 + 9.7 = -0.9 $\pm{3.5}$ km~s$^{-1}$. This is the
computed discrepancy for all measured photospheric wavelengths of 10~Lac 
from SWP LGAP observations referenced to the UV laboratory wavelengths
used in the Brandt et al. study. The error of $\pm{3.5}$ km~s$^{-1}$ 
is the sum added in quadrature of the values given for each of the
above four steps. This mean discrepancy appears as the first entry in
Table~\ref{tab3} and is the key single result of this paper. 

\subsubsection{ Alternative Atlases}
 
   Less precise comparisons can be made between {\it IUE SWP} camera 
data and the panchromatic {\it HST/STIS} atlas of Arcturus (Ayres, priv. 
comm.) and the {\it HST/GHRS} atlas of Procyon (Wood et al. 1996). 
Wood et al. measured Procyon atlas lines by detailed
centroid-fitting algorithms. The results were referred to an ``absolute"
(ground-based) scale, with a minor barycentric correction being applied for 
the star's orbital motion around a companion. These authors also noticed
a systematic trend of emission line velocities with excitation potential 
such that the velocity of the lower-excitation
features agrees with the ground-based scale within 1 km~s$^{-1}$.
The cross-correlation with the Arcturus atlas features was carried 
out for two available SWP echellograms and for eight echelle orders
at the long-wavelength end of this camera's range.
We measured a mean {\it IUE} - {\it STIS} velocity differences 
of -2.2 and +0.8 (both $\pm{5}$ km~s$^{-1}$), respectively, giving a mean 
of -0.7 $\pm{5}$ km~s$^{-1}$; see Table~\ref{tab3}. The comparison 
with the Procyon atlas could be made with only six observations.
Again, a flux deficit at short wavelengths and a general paucity of lines 
restricted our cross-correlations to echelle orders 72--76. The mean 
{\it IUE} - {\it GHRS} difference from this comparison was -4.9 
$\pm{5}$ km~s$^{-1}$. When corrected by +1.0 km~s$^{-1}$ (Irwin et al. 
1992), this result becomes -3.9 km~s$^{-1}$ (Table~\ref{tab3}). 
Errors of $\pm{5}$ km~s$^{-1}$ (taken from the comparison of {\it IUE/SWP}
data to the Procyon atlas) were applied to this comparison. In this case, the
data quality were higher, but there were fewer spectra in the sample. The
differences from comparisons with the two atlases, -3.9 $\pm{5}$ km~s$^{-1}$
and -0.7 $\pm{5}$ km~s$^{-1}$ are consistent with the value of -1 
km~s$^{-1}$ determined from the 10~Lac study.

\subsubsection{Comparison of 10 Lac with stars of similar spectral type}
\label{simtyp}

  Another check on the results of $\S$3.1 is to cross-correlate the 
{\it IUE} SWP/LGAP spectra of 10~Lac with stars of similar spectral type. 
Such comparisons 
check for target-centering errors of the reference star 10~Lac as well as 
propagation of errors in the {\it IUE/SWP} and {\it GHRS} calibrations
(i.e., a combination of errors arising from Steps 1 and 2).
Two stars which provide a ready comparison with 10~Lac are HD~93521 (O9 Vp) 
and HD~60753 (B2 III). We were able to cross-correlate the shifts between 
10~Lac and these two stars using nearly
all available spectra (at least 93 exposures in each case). To do so, we
again used echelle
orders $m$ = 70-119, except for $m$ = 103, 106, and 109, which are 
dominated by interstellar features. Correcting for differences between the 
optically-determined heliocentric radial velocities for the two stars (Wilson 
1953), we determined apparent radial velocity differences, in the sense 
``HD-star" versus 10~Lac, of +1.8 km~s$^{-1}$ $\pm{3}$ km~s$^{-1}$ and
-0.5 km~s$^{-1}$ $\pm{3}$ km~s$^{-1}$, respectively (see Table~\ref{tab3}). 
The errors listed are given by assuming 
that the errors inherent in each of the steps described above for 10~Lac are
similar for these stars and that they add in quadrature.

\subsection{Analysis of LWP Camera Data }
\label{long}

   Although the {\it GHRS} atlas of 10 Lac offers an excellent reference
template for the {\it IUE SWP} camera, there is no spectral atlas
of a nonvariable star which covers the long-wavelength camera range 
($\lambda\lambda$2000--3000) region and has a comparable quality. 
Thus, we had to resort to alternative strategies to evaluate zero-point 
errors in long-wavelength camera data. Although it is possible in 
principle to cross-correlate echelle orders for {\it SWP} and 
long-wavelength camera data over their overlapping wavelength range, 
practical difficulties intervene when carrying out this operation. One of 
these is that a loss of camera sensitivity in these regions causes a steep 
gradient in signal-to-noise ratio along the spectrum. In order to compare 
spectral lines of comparable quality, this constraint further limits the 
already narrow wavelength interval. In addition, spectra of even cool stars do 
not clearly exhibit many features. Despite these limitations, we were able to 
compute reliable cross-correlation shifts for the sharp-lined stars 10~Lac 
and $\tau$~Sco for wavelengths ($\lambda$1949--1959) in common with the two 
cameras, namely order $m$ = 71 for {\it SWP} and $m$ = 118 for {\it LWP/LWR}. 
For a comparison with the {\it LWR} camera, we find apparent radial velocity 
differences, RV$_{SWP}$ - RV$_{LWR}$ =  -0.1 and -3.0 km~s$^{-1}$ for 10~Lac 
and $\tau$~Sco, respectively. The corresponding results with the {\it LWP} 
camera are RV$_{SWP}$ - RV$_{LWP}$ = +1.1 and +5.4 km~s$^{-1}$, respectively.
As usual, these values refer to means of the ensembles of 
many {\it LGAP} spectra for each star. We estimate the errors in these 
comparisons to be about $\pm{4.4}$ km~s$^{-1}$ by adding in quadrature the 
typical order-to-order fluctuations  ($\pm{3}$  km~s$^{-1}$) with the 
average velocity difference ($\pm{3.2}$ km~s$^{-1}$) between large- and 
small-apertures. Since the derived zero-point differences are 
comparable with this r.m.s., our comparison suggests that there are no 
significant wavelength differences in the $\lambda\lambda$1950--2050 
region between the {\it SWP} and long-wavelength camera data. 

  An alternative method of evaluating mean radial velocity errors for the
long-wavelength errors is to cross-correlate the fluxes of {\it IUE} spectra
with those of {\it HST} atlases of Procyon and Arcturus. Note that the 
suitability of this technique relies upon our previously tying the 
zero-points of these atlases to the 10~Lac atlas via {\it IUE} data. 
This method is reliable in principle, 
but in practice it has the drawback that rather few {\it IUE} observations
are available for these particular stars. In the case of
{\it LWR} camera for Arcturus, we were able to cross-correlate 9 orders in 
just three spectra and found a difference, RV$_{LWR}$ - RV$_{STIS}$, of +2.7 
$\pm{4}$ km~s$^{-1}$. These quoted errors and others
quoted below include both the r.m.s. velocities for individual orders and 
the r.m.s. arising from the small numbers of observations. 
We cross-correlated seven available spectra with the Procyon atlas and 
found a difference of -3.5 $\pm{6}$ km~s$^{-1}$. A comparison between the 
Arcturus atlas and the {\it LWP} datasets is not possible because the 
{\it IUE} observations were made with the star image placed in various 
``nonstandard" positions in the large aperture. 
For the Procyon atlas (3 orders; 4 {\it LWP} spectra) the difference, 
RV$_{LWP}$ - RV$_{GHRS}$, is +7.9 $\pm{6}$ km~s$^{-1}$. 
However, this difference is noticeably affected by a wavelength 
shift of one spectrum (LWP~13112) with respect to the three.
Although we could find no {\it a priori} reason to exclude the LWP~13112 
spectrum from our analysis, its omission would result in an RV$_{LWP}$ - 
RV$_{GHRS}$ difference of +2.4 km~s$^{-1}$. This would bring the {\it LWP} 
scale fully into agreement with the essentially null difference found in 
the preceding paragraph. Additionally, we compared
the mean zero-point difference between the {\it LWP} and {\it LWR} cameras 
by cross-correlating many spectra in 59 orders of six stars. These results
are summarized in Table~\ref{tab4}. The entries in 
this table give a mean difference of only +0.4 $\pm{3}$ km~s$^{-1}$, which 
is shown as the last entry in Table~\ref{tab3}. This agreement suggests again 
that the zero-points of these two cameras agree to within errors of about
$\pm{5--6}$ km~s$^{-1}$.

To summarize all these comparisons, the radial velocity differences of the 
three {\it IUE} cameras, {\it inter alia}, are zero to within the 
measurement errors.

\subsection{Radial Velocity Zero-Point Errors: Dependence on Wavelength }
\label{rvsys}

  Because wavelengths were measured individually for more than 700 lines in 
the 10 Lac atlas, it was possible to search for wavelength errors as a function 
of wavelength in both the {\it GHRS} and {\it IUE/SWP} data for 10 Lac.
Figure~\ref{ms1fig12} depicts (small crosses) 
the differences between the laboratory lines
referred to in the Brandt et al. (1998) study and the measurements they made 
from their atlas lines (these differences have been corrected for the radial 
velocity of the star). We also plot as a solid line the cross-correlation 
results between the {\it GHRS} atlas and ensemble of {\it SWP} {\it IUE} 
spectra at each {\it SWP}-order bin (cf. Fig.~\ref{ms1fig9}).
This line shows mean fluctuations of $\pm{3}$ km~s$^{-1}$. Its
excursions from zero are always less than $\pm{3}$ km~s$^{-1}$. From
these results we conclude that there are no significant differences between 
the wavelengths of the {\it GHRS} and {\it SWP} camera. To some degree, this
is not unexpected because both calibrations were made using 
the same line list. Even so, the calibrations were derived
using different algorithms. In particular, the {\it GHRS} wavelength 
solutions were obtained by a block-grouping of lines in the entire spectral
domain (Lindler 1993) whereas the {\it IUE} solutions were performed for
individual echelle orders for wavelengths greater than $\lambda$1350. 
 
  Figure~\ref{ms1fig12} shows no trend in the mean velocities
for echelle orders grouped above and below the $\lambda$1350 demarcation. 
We can expect that both random and systematic errors of the more current 
Pt wavelengths 
are smaller than the older, heterogeneous sources. For example, one may
attribute the increased scatter of points in this figure below $\lambda$1500
to random errors from the older laboratory studies. Similarly, any
systematic errors in the older laboratory data would show up 
as trends away from both the zero-level and the solid line. 
The dashed line in the plot shows evidence of possible systematics in the 
laboratory data in the wavelength range $\lambda\lambda$1650--1750 and 
$\lambda\lambda$1190--1250. Detailed inspection of the line list from 
which the {\it GHRS} wavelengths are drawn suggests that such descrepancies 
are not limited to one type of ion. For example, several ions of both light 
and Fe-like ions are represented in these wavelength regions.

\subsection{Comparison with Other Studies }

  Two papers (Gonz\'alez-Riestra et al. 2000, Holberg, Barstow, \& Sion 1998) 
have reported that {\it NEWSIPS}-processed, high-dispersion 
{\it SWP} images have negative zero-points errors relative to other standards. 
Both rely upon measurements of centroids of individual interstellar lines, 
generally in spectra of OB stars. Gonz\'alez-Riestra et al. reported 
differences of -17.7$\pm{3.7}$ km~s$^{-1}$ for the {\it SWP} camera with 
respect to ground-based measurements of interstellar lines in the optical 
spectra of the same stars. A smaller discrepancy was found for small-aperture,
long-wavelength camera data. As a consequence, a correction of +17.7 
km~s$^{-1}$ was applied to all {\it SWP} high-dispersion spectra in the 
{\it IUE Newly Extracted Spectra} (``INES") system 
(Gonz\'alez-Riestra et al. 2000).

  Upon investigation we are unable to support the conclusion by 
Gonz\'alez-Riestra et al. that a significant difference in radial velocities 
exists for {\it NEWSIPS SWP} spectra. We have attempted without success to 
reconcile the differences between their results and ours. These authors kindly
provided us with program data lists, so we were able to compare 
cross-correlation shifts from their samples with those from entire sets of 
obervations of these stars in the archives. A comparison of the results 
showed no noticeable difference between the two samples of observations, 
and thus one must look elsewhere to resolve the difference.
Next, we cross-correlated the same interstellar lines
used in the Gonz\'alez-Riestra et al. analysis for in a sample of 145 
large-aperture {\it SWP} spectra of 10~Lac with the interstellar lines 
in the Brandt et al. (1998) {\it GHRS} atlas. For the three sets of lines 
they measured (Si~II $\lambda$1259--60, O~I $\lambda$1302, Si~II $\lambda$1304,
and C~III $\lambda$1335--6), we found a difference RV$_{SWP}$ - RV$_{GHRS}$ =
-0.9 $\pm{1.5}$ km~s$^{-1}$. Thus, there is essentially no difference 
between the systemic velocities of these interstellar lines and the 
photospheric lines in the {\it IUE} spectrum for these two instruments. 
Gonz\'alez-Riestra et al. did not measure {\it SWP} observations of 10~Lac, 
but their measurements of lines in HD~93521 and HD~60753 spectra suggest that 
the wavelengths in these stars typify the zero-point offsets they found for 
their total sample of six stars: 
-18.8 km~s$^{-1}$ for HD~93521 and -12.9 km~s$^{-1}$ for HD~60753.
In $\S$\ref{simtyp} we reported that the relative velocities of the 
{\it photospheric} lines for these three stars were in agreement with
ground-based studies, so in this important respect all our comparisons 
are self-consistent (as are those of Gonz\'alez-Riestra et al.).

  Since we can find no other discrepancy among our results with respect
to the atlas results or with respect to photospheric lines of the stars
in our common sample, we are left with the possibility that the measurement 
positions of the interstellar lines by Gonz\'alez-Riestra et al. are 
misleading for some subtle reason. 
One possibility is that the multi-component nature
of the interstellar clouds toward these stars gives different centroid
velocities for some groups of lines than for others. In particular, differing 
degrees of saturation among interstellar lines at low spectral resolutions 
can give rise to rather large centroid velocities differences among various
interstellar line complexes in the spectrum of a given star. For example,
differences between the $\lambda$1304 and $\lambda$1335--6 complexes are
quite large in the spectrum of HD~93521, as was indeed noted by 
Gonz\'alez-Riestra et al. themselves.

   In $\S$\ref{iuenew} we noted a very similar time-dependence of apparent 
velocities between {\it IUESIPS} and {\it NEWSIPS} spectra with respect to the
dependence noted by Holberg, Sion, \& Barstow (1998). These authors also found 
a mean difference of -8.3 km~s$^{-1}$ $\pm{1.4}$ between {\it NEWSIPS }-derived
and ground-based radial velocities for four white dwarfs. 
This finding appears to contradict the conclusion from our work on 10~Lac 
and other bright B stars ($\S$\ref{swpan}) that the velocity zero-point 
for {\it SWP} {\it NEWSIPS} is correct within -1 km~s$^{-1}$.
We have attempted to track down the source of this discrepancy in a number
of ways. First, in $\S$\ref{aperr} we discussed the possibility that the mode 
of centering on bright stars and faint stars could lead to different results 
(see Table~\ref{tab1}). Yet, we found no such effect. We may also consult
Fig.~\ref{ms1fig4} for clues of an apparent velocity difference between
white dwarfs and bright OB stars. The figure
implies that the mean of the white dwarf velocity offset should agree 
within about $\pm{1}$ km~s$^{-1}$ of our 10~Lac result. This statement holds 
true for the velocity average of the four white dwarfs selected by HSB. 
We have also cross-correlated the groups of interstellar lines considered by 
HSB in selected echelle orders. These comparisons produce a mean {\it IUESIPS} 
- {\it NEWSIPS} velocity shift of -1.3 $\pm{0.6}$ km~s$^{-1}$ relative
to differences derived from orders containing only photospheric lines.
Thus, there is nothing special or ``peculiar"
about the spatial mapping of the ISM lines in {\it NEWSIPS} that would
give results systematically different from an analysis of photospheric
features over the whole spectrum. 

  Unable to resolve a disagreement with HSB's results in this manner, we 
proceeded with another test. In their discussion of the apparent wavelength 
error, HSB compared their results with two earlier studies of interstellar 
lines in their white dwarf sample. The more accurate of these was a study by 
Lemoine et al. (1996; Small Science Aperture observations) of the ISM line
spectrum of the white dwarf G191-B2B. 
To make a comparison with {\it NEWSIPS} wavelengths, we acquired the Lemoine
et al. data from the the MAST archives. These observations were originally
made by substepping the grating to minimize fixed pattern detector noise, 
so we made corrections for these shifts by comparing emission line spectra 
obtained at nearly the same time for wavelength calibration 
(this comparison was made by measuring centroid line positions of the
two groups of spectra interactively). The differences we determined were
applied to the raw data in order to form high-quality, co-added spectra.
We next cross-correlated 13 {\it SWP LGAP} spectra of the star with the mean 
{\it GHRS} spectrum in narrow wavelength regions centered on 12 interstellar 
features studied by these Lemoine et al. These cross-correlations gave a mean 
shift, RV$_{NEWSIPS}$ - RV$_{GHRS}$, of -2.3 $\pm{3.0}$ km~s$^{-1}$, 
i.e. zero within the errors (see Table~\ref{tab3}). This finding is at
mild variance with the HSB prediction of -8.2 km~s$^{-1}$, but it is 
consistent with our other results. 

\section{Conclusions}

  Cross-correlation tools have been used to determine whether systematic 
trends in wavelength zero-points exist with respect to camera aperture,
time, focus, and THDA for high-dispersion {\it IUE/NEWSIPS} data for all 
three cameras. Not surprisingly, a major residual error source is in an 
implied inconsistency in target-centering/guiding among different observations,
some of which could arise from different thermal histories of the telescope.
Even so, the wavelength calibration of the {\it SWP} high-dispersion data 
is surprisingly good. In fact, it can rival the wavelength precision of the 
{\em Large} Aperture {\it GHRS} spectra for stars observed many times over 
several epochs and in consistent instrumental conditions. Even over 10--15 
years, we found evidence of significant velocity trends for only one or two 
stars ($\S$\ref{timetr}). At most these trends could amount to $\pm${3--5} 
km~s$^{-1}$ over this duration, and they may actually be smaller. We have 
also found that dedicated time-series observations of bright stars can 
substantially reduce apparent velocity errors. Yet, even for such data 
strings, changes in the spacecraft environment can produce spurious 
variations of $\pm{3}$ km~s$^{-1}$ over a timescale of a few hours.
It is not clear that series of long-wavelength spectra exist in
{\it IUE} database to detect such effects, but we expect that they 
are present nonetheless.

  Using the {\it GHRS} atlas of 10 Lac as an secondary standard, we have
found that the apparent radial velocity determined from high-dispersion images 
of the {\it IUE} {\it SWP} camera is zero, within the errors of measurement,
with respect to the ultraviolet wavelength system 
defined by contemporary laboratory catalogs (Brandt et al. 1997).
This result has been 
checked by lower-accuracy comparisons such as the difference between apparent
radial velocities from {\it SWP} echellograms of two other presumed constant
early-type stars with similar spectral type, HD~93521 and HD~60753, and also 
by comparing wavelengths of several interstellar lines with those measured 
by the {\it GHRS}. Additionally, and for all three {\it IUE} cameras, we have 
confirmed the accuracy of the zero-points with more limited sets of 
lines with the {\it GHRS} and {\it STIS} atlases of Procyon and Arcturus, 
respectively. We have compared the zero-points of the two {\it IUE} 
long-wavelength cameras by means of the Arcturus and Procyon atlases and 
also by compared wavelengths within a common narrow wavelength order among 
the three cameras. Neither of these comparisons led to a siginificant radial
velocity offset between the {\it IUE} and {\it HST} instruments. The  
long-wavelength camera results are also in agreement with the ground-based 
velocity of Procyon, but less precisely so. Our estimate of the accuracy of
in the zero-point error for the {\it LWP} and {\it LWR} cameras is itself 
imprecise, but judging from its consistency with the {\it Arcturus} and 
{\it Procyon} atlas wavelengths, it is probably no worse than $\pm{5}$ 
km~s$^{-1}$. Our findings are in apparent disagreement with those of 
Gonz\'alez-Riestra et al. (2000). Their results led to shifts in the 
wavelength scales of the {\it SWP} high-dispersion and {\it LWP/LWR} 
low-dispersion data for the {\it INES} ({\it IUE Newly Extracted Spectra}) 
data product produced by the {\it European Space Agency}.

  We have found no noticeable wavelength-dependent differences in zero-point
between {\it IUE SWP} camera data for 10~Lac and the {\it GHRS} atlas of
this star. 
However, note that this finding does not address possible errors within a
particular order or measurements of a common feature appearing in adjacent 
orders, such as the Mg~II h line (see Gon\'zalez-Riestra et al. 1998). 
Although we have not examined the data for this purpose, we have noticed 
occasional mismatches of 5--10 km~s$^{-1}$ from one end of an echelle order to
another compared to theoretical spectral line syntheses. Such errors probably
originate from an inadequacy of the calibration algorithms to determine 
dispersion nonlinearities accurately because only a few lines were available 
in individual orders of {\it WAVECAL} echellograms.

  Finally, because the {\it GHRS} atlas of 10~Lac has a broad 
wavelength coverage and shows good correspondence in its wavelengths
compared to laboratory data, we recommend the use of this atlas as a 
reference standard for comparison 
of astronomical mid-UV wavelengths with data from other missions.

  The author gained valuable technical information presented herein while 
serving on the staff of the {\it IUE} Project at the {\it Goddard Space Flight
Center} (NASA).
It is our pleasure to acknowledge informative technical discussions with Dr. 
Catherine Imhoff,  Mr. Randall Thompson, and Mrs. Karen Levay concerning 
questions of {\it IUE} operations practices and detailed contents of the 
{\it IUE} database. We also thank Dr. Rosario Gonz\'alez-Riestra for 
providing us with a list of spectra used in her team's investigation of
{\it NEWSIPS} data properties.
We acknowledge helpful conversations with Drs. Richard Robinson, Charles 
Proffitt, and Mr. Don Lindler on the calibration of {\it GHRS}
wavelengths, and Dr. Gill Nave for comments on the utility of various 
laboratory wavelength sources. We are also indebted to to Drs. Martin Snow 
and Brian Wood for providing their atlas data to MAST, and to Dr. Tom Ayres 
for providing it before publication. We are grateful to Drs. Rosario 
Gonzalez-Riestra and Jay Holberg for providing unpublished data from their
work. We wish to thank Dr. Marc Postman for his patience. 

\newpage

\begin{references}

\reference{} Ayres, T. R., Brown, A., Harper, G. M. et al. 1999, 
BAAS, 194, 6701A

\reference{} Barstow, M. A., Holberg, J. B., Bruhweiler, F. C., Collins, J.
1996, \aj, 111, 2361 

\reference{} Brandt, J. C., S. R. Heap, et al. 1998, \aj, 116, 941

\reference{} Fullerton, A. W. 1990, Ph. D. Dissertation, Univ. of Toronto

\reference{} Garhart, M. P., Smith, M. A., Levay, K. L, Thompson, R. 
W., \& Turnrose, B. E. 1997, NEWSIPS Manual, IUE NASA Newsletter No. 57

\reference{} Garhart, M. P., Smith, M. A., Levay, K. L, Thompson, R.
W., \& Turnrose, B. E. 1997, NEWSIPS Manual, IUE NASA Newsletter No. 57

\reference{} Gonz\'alez-Riestra, R., Cassatella, A., Solano, E., Altamore,
A., \& Wamsteker, W. 2000, Astr. Ap. Suppl, 141, 343

\reference{} Grayzeck, E. J., \& Kerr, F. J. 1974, \aj, 79, 368

\reference{} Hobbs, L. M. 1969, \apj, 157, 135

\reference{} Holberg, J. B., Barstow, M. A., \& Sion, E. M. 1998, 
\apjs, 119, 207

\reference{} Irwin, A. W., Fletcher, J. M., Yang, S. L. S., Walker, G. A. H.,
\& Goodenough, C. 1992, \pasp, 104, 489


\reference{} Leomoine, M., Vidal-Madjar, A., Bertin, P., Ferlet, R., Gry,
C., \& Lallement, R. 1996, Astr. Ap., 308, 601L

\reference{} Lindler, D. 1993, in Calibrating Hubble Space Telescope,
ed. J. C. Blades \& S. J. Osmer (Baltimore: STScI), p. 278


\reference{} Reader, J., Acquista, A., Sansonetti, C. J., \& Sansoneti,
J. E. 1990, \apjs, 72, 831

\reference{} Robinson, R. D. 2000, priv. commun.

\reference{} Rogerson, J. B., Jr., \& Upson, W. L. II 1977, \apjs, 35, 37

\reference{} Stokes, G. M. 1978, \apjs, 36, 115

\reference{} Walborn, N. R. 1972, \aj, 77, 312

\reference{} Wilson, R. E. 1953, General Catalogue of Stellar Radial 
Velocities (Washington: Carnegie Inst).

\reference{} Wood, B. E., Harper, G. M., Linsky, J. L., Dempsey, R. C.
1996, \apj, 458, 761

\end {references}

\setlength{\textwidth}{183mm}
\setlength{\evensidemargin}{22mm}

\begin{table}[htb]
\begin{center}
\caption{\label{tab1} Large Minus Small Aperture Zero-Point Differences (km~s$^{
-1}$) }

\centerline{~}

\begin{tabular}{r|rrrrrr} \hline\hline
SWP Camera: &   &  &  &  &  &  \\ \hline
Bright Stars:   & 10 Lac & $\tau$ Sco & $\zeta$ Oph & $\eta$ UMa & $\lambda$
Lep & $\zeta$ Cas  \\
LGAP - SMAP  & 0.4  &  -3.5  & -0.6    & +0.2 & +0.2    &  -1.7 \\
No. LGAP: &  145  & 74 & 59 & 48 & 40 & 54 \\
No. SMAP: &  6  & 33 & 15 & 12 & 10 & 12 \\
\hline
Faint Stars:   & RR Tel & Sirius B  & BD+75$^{o}$325 & BD+28$^{o}$4211 & HD\,93521  & 
HD\,60753 \\
LGAP - SMAP:   & -2.5 & -2.3 & -4.7 & 2.4 & 2.1 & 0.7   \\
No. LGAP: &  46  & 3 & 86 & 76 & 99 & 82 \\
No. SMAP: &  5  & 1 & 5 & 2 & 14 & 1 \\
\hline\hline
LWP Camera: & $\tau$ Sco  & $\eta$ UMa  & HD\,93521  &  &  &  \\ \hline
LGAP - SMAP  & -2.5  &  -1.1  & 0.1    &  &     &   \\
No. LGAP: &  75  & 156 & 20  &  &  &  \\
No. SMAP: &  15  & 4 & 2 &  &  &  \\
\hline\hline
LWR Camera: & 10 Lac & $\tau$ Sco  & $\zeta$ Oph & RR Tel &  &  \\ \hline
LGAP - SMAP  & 1.9  &  -1.0  & -4.3    & +4.3 & &  \\
No. LGAP: &  3  & 24 & 11  & 15  &  &  \\
No. SMAP: &  4  & 14 & 7 & 3 &  &  \\
\hline\hline

\end{tabular}
\end{center}
\end{table}

\begin{table}[htb]
\begin{center}
\caption{\label{tab2} Linear Regression Velocity Slopes with Time
for {\it LGAP } Data (km~s$^{-1}$ yr$^{-1}$) }

\centerline{~}

\begin{tabular}{r|rrrrrrrr} \hline\hline
Star/Cam.:   & 10 Lac & $\tau$ Sco & $\zeta$ Oph & $\eta$ UMa & $\lambda$ Lep &
$\zeta$ Cas & HD~60753 & +75$^{o}$325 \\
SWP  &   &  &  &  &  &  & & \\ \hline
Slope:  & -0.01  &  -0.05  & -0.29    & -0.49 & -0.13   & -0.03 & -0.53
& -0.34  \\
No. $\sigma$: & ($<$1)    & ($<$1)    & 2.1 &  (1.5) &  (1.0) & ($<$1)
& $>$4 & 4 \\
\hline
  &   &  &  &  &  &  \\ \hline
LWP &   &  &  &  &  &  &  &  \\ \hline
   & 10 Lac & $\tau$ Sco & $\eta$ UMa & +28$^{o}$4211 & $\zeta$~Cas &
+75$^{o}$325 & RR Tel & HD\,93521 \\
Slope:  &  -0.44 & -0.10 & 0.01  & 0.04 & 0.28 & -0.03 & 0.05 & .03 \\
No. $\sigma$: & ($<$1) & ($<$1) & ($<$1) & ($<$1)  & $<$1 & $<$1 & $<$1
& $<$1 \\
\hline
LWR &   &  &  &  &  &  &  &  \\ \hline
No. $\sigma$: & ($<$1)    & ($<$1)    & 3.1 &  (1.5) &  (1.0) & ($<$1)
 &   &   \\
Star:   & $\tau$ Sco & $\zeta$ Oph & $\zeta$ Cas & +28$^{o}$4211 &  &  & & \\
Slope:  & -.57  &  -0.55  & 0.28 & -.28  & -0.34  & -1.7 &  &  \\
No. $\sigma$: & ($<$1) & ($<$1)  & ($<$1) & ($<$1)  &  &  &  & \\
\hline\hline

\end{tabular}
\end{center}
\end{table}

\begin{table}[htb]
\begin{center}
\caption{\label{tab3} Radial Velocity Zero-Point Results (``Other" - 
IUE$_{SMAP}$) }
\centerline{~}
\begin{tabular}{ccr} \hline

Camera &  Comparison Made & Mean RV Diff \\ \hline \\
{\em SWP} & & \\
 & UV lab via GHRS atlas: 10 Lac ~(main result) & -0.9 $\pm{3.5}$ km s$^{-1}$ \\
  &   &   \\
 & Procyon atlas (GHRS, em. lines) &  -3.9 $\pm{5}$ km s$^{-1}$ \\
& Arcturus atlas (STIS, continuum) & -0.7 $\pm{5}$ km s$^{-1}$ \\
& Cross-corr. HD 93521  with 10 Lac (IUE) & +1.8 $\pm{3}$ km s$^{-1}$ \\
& Cross-corr. HD 60753 with 10 Lac (IUE) & -0.5 $\pm{3}$ km s$^{-1}$ \\
  &   &   \\
 & Cross-corr. with GHRS of ISM lines (G191-B2B) & -2.3 $\pm{3}$ km s$^{-1}$ \\
{\em LWR} & & \\
 & Procyon atlas (GHRS) &  -3.5 $\pm{6}$ km s$^{-1}$  \\
 & Arcturus atlas (STIS) & +2.7 $\pm{3}$ km s$^{-1}$  \\
  &   &   \\
{\em LWP} & & \\
 & Procyon atlas (GHRS) &  +7.9 $\pm{6}$ km s$^{-1}$  \\
LWR - LWP: & X-corr. of 6 stars & +0.4 $\pm{3}$ km s$^{-1}$  \\
Long \& short-$\lambda$ cams.  & & \\
 & SWP - LWP: & +3.3 $\pm{4}$ km s$^{-1}$  \\
 & SWP - LWR: & -1.5 $\pm{4}$ km s$^{-1}$  \\
\end{tabular}
\end{center}
\end{table}

\newpage

\setlength{\textwidth}{183mm}
\setlength{\evensidemargin}{22mm}

\begin{table}[htb]
\begin{center}
\caption{\label{tab4} Radial Velocity Differences between Long-Wavelength
Cameras (LWR - LWP) }
\centerline{~}

\begin{tabular}{ccr|ccr} \hline \hline
Star &  Mean RV      & Number   & Star      & Mean RV       & Number \\ 
     & Difference    & Spectra  &           & Difference    & Spectra \\ 
     & (km~s$^{-1}$) &          &           & (km~s$^{-1}$) &  \\ \hline \\ 
10 Lac & 1.0 & 3, 13 & RR Tel & -3.5 & 21, 14 \\
BD +75$^{o}$325 & 3.4 & 6, 67  & $\tau$ Sco & 3.3 & 24, 75 \\ 
BD +28$^{o}$4211 & -3.1 & 13, 45 & HD 11636  & 1.3 & 12, 2 \\
\hline \\
\end{tabular}
\end{center}
\end{table}

\newpage
\clearpage
 
\begin{figure}
\vspace*{.5in}
\caption{
Cross-correlation shifts of 415 {\it WAVECAL}
{\it SWP} echellograms during the
lifetime of the {\it IUE} for
orders at the opposite ends of the camera, centered at 1168~\AA~ 
and 1969~\AA.}
\label{ms1fig1}
\end{figure}

\begin{figure}
\caption{
Velocity shifts
of {\it SWP} camera WAVECAL echellograms during the {\it IUE}
lifetime. The solid line is a cubic polynomial fit (zero-point arbitrary)
which shows the (incomplete) extent to which time dependences were removed 
from the {\it WAVECAL} wavelength zero points.
Note that at certain epochs the difference between an local mean and the
smooth curve may amount to $\approx$ 2 km~s$^{-1}$. Such residuals are not
removed in the wavelength calibration of {\it NEWSIPS}.
}
\label{ms1fig2}
\end{figure}

\begin{figure}
\caption{
Wavelength-integrated cross-correlation shifts of {\it WAVECAL} echellograms
of the two {\it IUE} long-wavelength cameras
during {\it IUE} lifetime. Shifts represent wavelength
zero-point differences and are expressed as velocities. Solid line depicts
the fit with a linear ({\it LWP}) and cubic ({\it LWR}) function.
}
\label{ms1fig3}
\end{figure}

\begin{figure}
\caption{
Cross-correlation shift differences in the sense {\it IUESIPS} - {\it NEWSIPS}
with time. Displayed are shifts for echellograms observed for the
velocity standards (as labeled). The white dwarfs are those 
studied by Holberg et al. (1998). Vertical dotted lines
indicate the times in which new wavelength calibration constants were applied
to the {\it IUESIPS} processing software according to Thompson et al. (1981),
Thompson et al. 1983, Thompson \& Turnrose, 1983, Gass \& Thompson 1985, \&
Garhart 1993).
}
\label{ms1fig4}
\end{figure}

\begin{figure}
\caption{
Apparent velocity trends with time: (a) HD 60753 and (b) BD+75\,325.
All data were obtained through the large aperture.}
\label{ms1fig5}
\end{figure}

\begin{figure}
\caption{Cross-correlation shifts of velocity, telescope focus, and THDA
temperature for an intensive time series of {\it SWP} high-dispersion
observations of $\alpha$~Eridani. Dotted and dashed lines (shown for
reference only) represent the x- and y-components of the satellite's orbital
motion (``x" is directed toward the Sun; ``y" its perpendicular in the
Earth's equatorial plane). 
THDA is in degrees Centigrade; focus is in instrumental units.
}
\label{ms1fig6}
\end{figure}

\begin{figure}
\caption{Cross-correlation shifts of velocity, telescope focus, and THDA
temperature for an intensive time series of {\it SWP} high-dispersion
observations of $\epsilon$~Persei. For reference only: a cross-correlation
between the apparent velocities and the x-component of the (diurnal) orbital 
velocity shows that the velocities show a mean lag of 0.15 cycle behind the 
x-component.
}
\label{ms1fig7}
\end{figure}

\begin{figure}
\caption{Velocity differences versus {\it IUE} telescope focus for
echellograms of the data depicted in Fig.\ref{ms1fig7}.}
\label{ms1fig8}
\end{figure}

\begin{figure}
\caption{
Mean shifts of wavelength zero-points as a
function of wavelength extracted from a large number of {\it IUE} echellograms
for 16 stars (10~Lac, $\tau$~Sco, $\zeta$~Cas, $\zeta$~Oph, $\eta$~UMa,
$\lambda$~Lep and the white dwarfs BD+75$^{o}$325, BD+28$^{o}$4211, Sirius~B, WD0005+511,
NGC246, REJ0623-3, EG102, Wolf~1346, GD~394, Gl191-B2B). These shifts were
obtained by cross-correlating wavelengths in each high-dipersion echelle order
for a number of test images of a star against a reference echellogram. Belts
of $\pm{1\sigma}$ around the mean shifts are drawn from the distribution
of the shifts for the individual stars.}
\label{ms1fig9}
\end{figure}

\begin{figure}
\caption{Large-aperture -- Small-aperture {\it SWP} apparent radial velocities 
of 10 Lac during the lifetime of the {\it IUE}. The sample comprises 
145 LGAP images and is referred to the average of 6 SMAP images.
The large scatter among velocities at late epochs is likely due to acquisition
of the star at a different position in the {\it FES}.
This diagram shows that the
instrumental {\it SWP} velocities of 10 Lac exhibit no significant trend with
time. Error bars are taken from Fig.~11.}
\label{ms1fig10}
\end{figure}

\begin{figure}
\caption{Velocity histogram for radial velocities for 10~Lac taken from 
Figure 10. This distribution is essentially bimodal, reflecting 
two sets of target centerings for late epoch observations.}
\label{ms1fig11}
\end{figure}

\begin{figure}
\caption{Differences of {\it GHRS} - {\it IUE} (jagged line)  and {\it GHRS} - 
{\it laboratory} (points) wavelengths for photospheric lines of 10 Lac as a 
function of wavelength. The {\it GHRS} - {\it IUE} differences refer to
cross-correlation shifts between lines in the 10 Lac atlas and in a group
of 145 high-dispersion {\it SWP} echellograms. The {\it GHRS} - {\it Lab}
differences have been shifted by +16 km~s$^{-1}$ to account for the combined 
radial velocity of the star and an instrumental error of the {\it GHRS}
relative to the laboratory system. Error bars in the lower right are for the
jagged line and refer to the mean difference between adjacent {\it SWP} 
echelle orders.
}
\label{ms1fig12}
\end{figure}

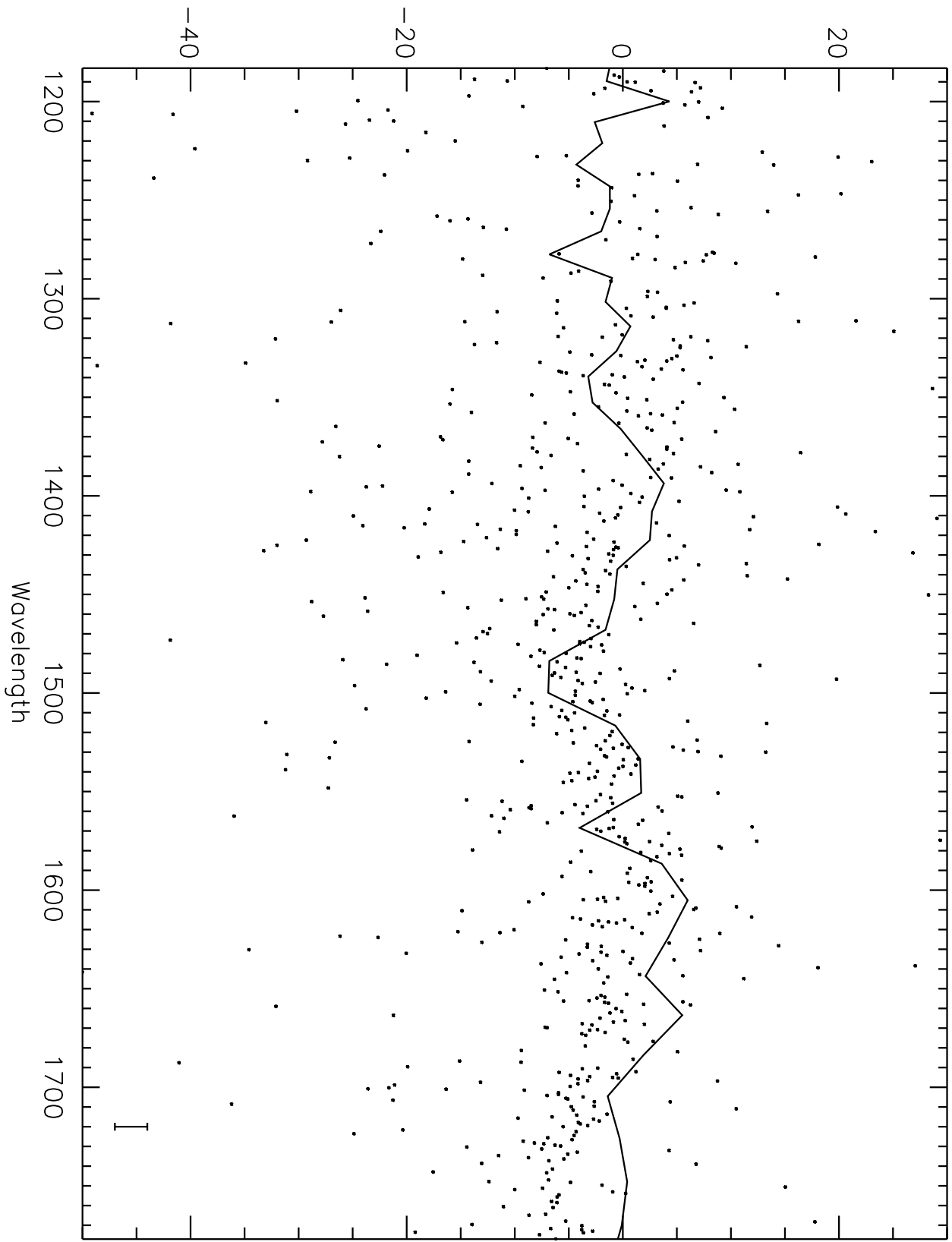
\end{document}